# Pandemic and disability: Challenges faced and role of technology

**Monnie Parida**  
mpiitkgp19@gmail.com

**Dr Manjira Sinha**  
manjira.sinha@cet.iitkgp.ac.in

**Indian Institute of Technology, Kharagpur, India-721302**

**Abstract.** The pandemic has affected every facet of human life. Apart from individual's psychological and mental health issues, the concern regarding mobility, access and communication with high risk infection is a challenging situation. People with disability are more likely vulnerable to infections. The new changes in our social lifestyle (social distancing, limiting touch) can profoundly impact the day today life of people with disability. In this paper, we will briefly discuss the situation faced by individuals with disabilities, some known remedies, and yet to be identified and curated technological remedies; the impact due to transition of special education toward online mode. Tips and tricks for better utilization of work from home concept by people with disabilities. Accessibility must be universal, accommodating all and encouraging inclusivity. As rightly said by Helen Keller, "The only thing worse than being blind is having sight but no vision"; subsequently, going by the demand of the time, we should contribute toward the universal design approach by supporting people with disabilities and commit to the changes required in disability care to reduce the impact of pandemic.

## 1. Introduction

International Classification of Functionality, Disability, and Health (ICF) prepared by WHO defined disability as "Impairments in body function and structure, participation, restrictions and activity limitations" [3]; overall, these are the components that form what the WHO calls 'Health Condition (disease or disorder). As we advance the change in definition [4] mentions, "Disability is the outcome of a complex relationship between an individual's health condition, personal factors, and the external factors that represent the circumstances in which the individual lives". Overall the above defination talks about the medical and social model of disability [6] which does not justify the right and equal opportunity for people with disability in different situations. While social model prioritizes the use of products, facilities, and services designed to remove accessibility barrier through society and environmental factor [3]. The medical model regards disability as a specific pathological issue restored through assistive technology and prosthetics [4, 11]. In this paper we suggest the non tragic view of disability - the Affirmative model [20] in combination with the social and medical model of disability. By saying non tragic, the focus is on positive understanding towards disability. Covid 19 pandemic is reasonably anticipated to have a higher impact for people with disability than people without disability [22]. Proceeding forward in this paper, we will have instances justifying the impact on people with disability. We are suggesting technological intervention and mass awareness about special needs with regards to pandemic. These measures can be further helpful for policy makers in similar future situations.

---

[1]GLAD Statement on COVID-19: Global action on Disability Network Statement on persons with disabilities in the COVID-19 outbreak and response, https://gladnetwork.net/sites/default/files/2020-04/GLAD- Statement-on-COVID-19 FINAL.pdf

We have around one billion persons with disabilities globally [1]. A survey carried out in India for 1067 persons (73% male, 27% female) shows around 73% of them facing challenges due to lockdown. Further, 201 people interviewed mentioned; 67% had no access to doorstep delivery while 22% had access to basics, 48% had no access to a state helpline, and 63% had not received the monetary support for people with disabilities announced by the Finance Ministry [2]. There is no single solution to the different sets of problems faced by person with Disability. According to [1] economic disparity, poor health, poverty, and lower educational attainment can amplify the distress situation for a person with disability during the pandemic. The absence of chronological treatment for people with disabilities can result in a higher chance of relapse of the previous stable state of their health [27]. Even access to caregivers through telemedicine is not an absolute support for a person with disability [9]. The care giver's skills and knowledge in the smooth handling of telehealth services are debatable [27]. The overall convenience to telehealth services for a person with disability is challenging, justifying the pandemics poor health outcome. The disruption in continuity of speech therapy and psychological treatment hinders the progress of educational attainment [27]. The scarcity of supportive employment effects in reduced quality of life for people with disabilities, creating an economic deficit situation for them [5]. Barriers to access health care can result in an unreasonable impact if due consideration is not given to the person with disability in the fight against COVID 19 [1]. Effective awareness and technological intervention are the keys to an inclusive environment involving people with disabilities [14].

This paper seeks to raise awareness and pave the way for new research ideas for a better living environment for people with disabilities in any situation similar to the pandemic. The disabilities Act identify disability as visual disability, low vision, leprosy-cured, auditory disability, loco-motor disability, mental retardation, and mental illness [1]. In the following sections, we have covered the above disabilities and their remedial measures. Education being an important part of remedial measures; therefore, we have also covered the overall effect of online education for people with disability. The Era of work from home can be better explained and learned through the eyes of people with disability who have been following the working from home option for decades. To support our approach toward better living environment during any similar situation, we have discussed the various successful implementations of remedial policies and measures followed around the world.

## 2. Challenges faced by people with visual disability during pandemic

For people with visual disability, the mode of communication and recognizing the object involves the sense of touch, feeling, and reacting to a situation [15]. In India, about 13 million people with different degree of visual disability are prone to covid 19 [15]. The reliance on physical contact and the need for communication with support persons; challenges the call for social distancing and inadvertently exposes the individual with visual disability to a contaminated surface [2]. That is, people with visual disability are more susceptible to get infected with covid 19. During lockdown, a distress situation is created due to severing of services and support network; serving as communication gateway [17]. As most people with visual disability belong to rural India with lower socio economic background, the morbidity and mortality is anticipated to be more significant if they are affected by covid 19 [15]. Due to lack of awareness and ignorance among people around us, the hurdles concerning pandemic have increased multiple times. The present scenario limits access to public health, hygiene information, and resources to implement good hygiene [2]. Even websites providing essential information is not always updated with W3CAG standards for effortless accessibility for people with disabilities [15].

---

[2] The Hindu, https://www.thehindu.com/news/national/persons-with-disability-face-severe-challenges-in-lockdown-report/article31643450.ece.

[3] Open letter to the UN Secretary General from the thematic group on Disaster Risk Reduction calling for the inclusion of PWDs during COVID19 responses, https://disabilityrightsfund.org/wp-content/uploads/TG-DRR-COVID-letter-v.1.0-230320-v2-.docx.

*Remedial approach*

As a remedial measure for access to public communication, alternative phone lines for people with visual disability can be a temporary option[12]. Smart phone application like "Be My Eyes" can be a convenient solution [28]. These smart phone applications can provide access to registered volunteers through audio calls and video calls for any emergency or other assistance [28]. We suggest using this application for communicating information regarding vaccination drive and hygiene routine. For outdoor mobility, a long cane can be used, asking the sighted person to hold the opposite edge of the cane to avoid direct body contact, thereby adhering to social distancing norms [15]. People with visual disabilities depend on voice as a human face for learning and communication. The innovative solution of talking pen and digital pen based on the multimedia print reader(MPR) used for hearing the text while reading by people with visual disabilities can be helpful not only at the time of pandemic but also in general. This innovative educational technology helps people with visual disabilities learn independently and is a blessing for teachers when not able to provide personal attention in multi-level teaching and learning environment [8]. Another cutting edge technology of using braille tactile connected with laptops during online classes is an empowering aspect for people with visual disability[6]. People with visual, hearing or intellectual impairments have difficulty accessing support assistance, aid, and messages to prevent COVID-19 as they are not provided in versatile and convenient formats. The information available through braille format in large prints is easily accessible to people with visual disability [9]. The assistive technology of screen reader can help the smooth access of Web content for people with visual disability [9]. Simultaneously, we should abstain from providing information through the online slideshow to people with visual disabilities as it is not compatible with a screen reader or magnification technology. Although videos have captions, captions in the comment section help people who cannot read fast and face difficulties with the text size. Provision of transcript for essential information is also beneficial.

**3. Challenges faced by people with auditory disability during pandemic**

People with auditory disabilities are stigmatized as slow cognitively even when the actual problem lies with the complication involved in transmitting knowledge and how communication is carried out [12]. People with an auditory disability do not acquire any language by hearing when spoken; most of them have parents who can hear; thus, they don't use Sign language from childhood. The studies show that part of the brain dealing with language degenerates by the third year of age if unutilized, and simultaneously the capacity to acquire language decreases over time[7]. It is crucial for the overall development of speech and language that there should not be any sensory and auditory stimuli loss. Without knowing sign language, the person cannot express and experience communication isolation, low skills, and stereotyping[7]. For a person with an auditory disability, impairment level depends on the amount of hearing loss. A person learns a language through memorizing, but the stress lies with compelling the person with an auditory disability to lip read and use speech. Still, the reality is they prefer Sign Language over any other language for communicating within a group of people with hearing disabilities [12]. The present pandemic era has created a problematic situation for people with auditory disabilities; they receive less information than they used to; Calling a helpline number and expressing themselves is a big challenge. A person with hearing impairment communicating in sign language does not necessarily mean the person they are talking to also understands sign language [7]. At times of emergency, this creates a life and death situation. Even denial of support services like sign language interpreter creates the communication gap. Some Indian sign languages for regional languages lack essential

---

[4]Pennsylvania Assistive Technology Foundation: Practical ideas for health and safety at home with a dis- ability during COVID-19,https://patf.us
[5]Novel corona virus; Advice for public, https://www.who.int/emergencies/diseases/novel-coronavirus-2019/advice-for-public

Information like they may not have a sign for the corona virus or quarantine. Lip reading is a critical part of sign language, but it is challenging to analyze when people wear masks, and not wearing a mask can create a health hazard not only for them but for people around [10]. Repair and maintenance of hearing aid is a concern with different phases of lockdown on our way to complete recovery from the pandemic.

*Remedial approach*

In addition to the increasing feeling of separation, depression, and stress [18], the further disappointment of not being capable of interacting over the telephone with medical specialists, and loved ones; leave the person with auditory disability disconnected from their support network. At this time, the captioning phone with an unprecedented demand for captioning is an empowering approach for communicating with medical specialists, keep in contact with friends and family [8]. As the lockdown has been extended widely around the Indian diaspora, a separate email address for a person with auditory disability may be a substitute to acquire and convey essential information[12]. Further remedial measures involve having sign language interpreters and access to skilled health and child protection personnel[9]. The use of a transparent mask for ease of lip reading is one of the many guidelines supported by WHO Moreover, using of a face shield instead of a mask is another viable option [10]. Printed formats or videos with text captioning and sign language, including subtitle/close-captioning, pictures with messages in sign language can be remedial solutions toward better information transmission among people with auditory disabilities [1]. The electronic media[12], government guidelines and information should be available in accessible formats like closed captioning, relay services, text messages, sign language interpreters, and easy to read language[11].

### 4. Challenges faced by people with loco motor, multiple and other disability

For any disability, if the person requires physical help, social distancing proves it to be a myth in the present situation of the COVID 19 pandemic [2]. The regular medical attention and medicine required is a luxury for people with disability. For a person with prosthetic limbs, it isn't easy to disinfect it with soap and water inadvertently, thereby exposing it to a contaminated surface. The need for the hour requires regular cleansing of touched surfaces like assistive tools, magnifying glasses, wheelchairs, crutches, etc. [7]. Disruption in health services that they routinely rely on can result in an elevated risk of serious illness [2]. The diverse need of a person with psychological disorders requires attention. Increased isolation exacerbates their condition [9]. The necessary medical facility for their caregiver is a concern that needs attention[1]. People with Down syndrome rely on images and videos for their daily chores. Therefore it isn't easy to follow the required hygiene regime to prevent corona virus without access to pictures and videos explaining the need to follow the hygiene rituals.[10] Autistic persons and people with psychosocial disabilities require care and understanding as they might not cope with strict confinement at home [14].

---

[6]Deccan chronicle: The touch of technology for blind kids, https://www.deccanchronicle.com/technology/in- other-news/120520/a-touch-of-technology-blind-kids-in-jk-to-get-laptops-to-link-to-bra.html

[7]National Association of Deaf 2013

[8]Pennsylvania Assistive Technology Foundation: Assistive technology spotlight; Captioned phones help with communication during quarantine, https://patf.us

[9]Unicef for every child: Remember COVID-19 response; Considerations for children and adults with disabilities, https://www.unicef.org/disabilities/files/COVID-19

[10]World Health Organization: Corona virus disease (COVID19) advice for the public, https://www.who.int/emergencies/diseases/novel-coronavirus-2019/advice-for-public

Information about covid 19 prevention and care must be available in accessible formats for people with intellectual disabilities. We can have image description for images and video descriptions for videos. Social media post with awareness and information about health and hygiene should be short and in plain language. Lengthy posts can be arranged in the comment section. Hash tags can be used in the comment section but should not be within the post. Additional guidelines and suggestive approach should be provided to people who cannot clean their hands regularly on their own and need access to water for the cleanliness regimen [7]. Even though vital, the quarantine centres are not disabled and elderly-friendly. We can still incorporate as much as possible for easy access to the disabled and elderly at this juncture. People with Down syndrome and their acquaintances can download images and guidance from the WHO website[10]. Helpful clues with pictures placed around the house as a reminder to wash hands and disinfect the mobile every time they enter their home are useful tricks toward preventing corona virus[4]. A noosed UV sanitiser with timers, blinker as a reminder to wash hands for 20 seconds helps follow the health hygiene [4]. People striving to cope with strict confinement at home can go for quick and careful outings during the day, which can be essential for them to cope with the situation[14]. The technology has given a priceless glimpse at something we may have missed otherwise for people with Down syndrome. As a silver lining to the quarantine situation, some find it easy to communicate through different kinds of assistive technology like Zoom, Face time, and snail-mail [17].

### 5. Online learning and people with disability

Online learning can present the opportunity for more inclusive education for a student with disability [25]. Our present education system should be remodeled to accommodate each childs need to establish inclusivity. The digitally supported learning model where the internet is used for teacher student interaction can be defined as an online mode of education [23]. Students with disabilities may be at risk of segregation from learning if online education plans are not feasible through assistive devices to support participation and aid education needs [9]. Just like other schools, the special schools are moving towards the online mode of education [24]. Students with disabilities often undergo physical, speech, mental health, and other therapies at schools, but as schools around the nation are closed, disabled students and their families are left without support [29].

The continuity of education ensures inclusivity for people with disability, considering education being an essential part of the remedial solution. The progress toward improved asynchronous learning leaves students with disabilities behind; for example, videos are often not captioned and many courses taught are designed without any accessibility support [19]. Students with hearing impairment may need changes about how to access audio material through transcription or sign language. People with visual disability depend on a screen reader, screen magnification software and visual aids [25]. Many students require customized tools, personalized assignments and individualized care to succeed. Educative technologies will be further convenient if they create the education context to produce a supportive, relaxed, and connecting learning environment [19].

---

[11]United Nation Human rights, Office of the High commission: COVID-19 Who is protecting the people with disabilities? https://www.ohchr.org

[12]The Albino Foundation: COVID 19 Disability inclusion emergency response by the albino foundation, Disability inclusion Nigeria, https://albinofoundation.org/covid-19-disability-inclusion-emergency- response-by-the-albino-foundation-disability-inclusion-nigeria

[13]Washington post: Disabled people have worked remotely for years, and they have got advice for you and your bosses, https://www.washingtonpost.com/lifestyle/wellness/disabled

[14]Disability rights fund: Creating accessible social media for those with deaf blindness https://disabilityrightsfund.org

At the same time, online education has proved beneficial for some students with disabilities. For example, going online might decrease the demand for disclosure and improve adaptability if instructors are accommodating. It reduces the barriers for students with mobility-related impairments [19]. With the onset of the online teaching and learning approach, it has proved to be empowering for individuals with special needs who prefer less distraction and less pressure[16]. Online learning provides natural flexibility in terms of time and place[16]. Independence is increased through the availability of online learning resources. Repetitions and patience are the special educator's mandatory requirements to aid the various necessities of people with disabilities [26]. We propose information channels for people with disabilities. As many students with disabilities do not have access to schools, any relevant information circulated through an educational institute may not reach children with disabilities. In extension to this, persons with disabilities may have less access to social media and other technological platforms [9]. During online learning, it has to be assured that distance education platforms are secure and convenient for children with disabilities; educators are equipped to assist remotely to ensure continuity of education [25]. Considering individual care and guidance for children at home, their mental well-being [9] are some of the measures to be initiated while supporting the online learning for people with disabilities.

## 6. The ray of hope

Barbarin, Taryn Williams, managing director of the Poverty to Prosperity program at the Centre for American Progress, isn't satisfied, organizations will abruptly allow extensive work from home opportunities after the corona virus crisis settles down. Though she does believe the change will commence to "redoing the way we think about being in public and what community looks like," [13]. The idea of working remotely instead of office environment best describes the concept of work from home [30]. Before pandemic, the type of occupation was the vital determinant to decide the feasibility of work from home [30]. Some profession has a higher acceptance of working remotely due to their nature of work [31]. During the pandemic, social distancing and stay at home policies have not only played a critical role in restricting the spread of the virus. It has opened vast opportunity and acceptance for the work from home concept. The advantage of work from home during pandemic reduces the negative employment impact [30]. Many people can still retain their jobs by working from home then losing their livelihood suddenly. People who cannot work from home may negatively affect their well being and experience discomfort [13]. Disable individuals has been working remotely for decades now [16]. Their experience is the key to manage the process of working remotely more efficiently. Thereby presenting the society with successful prototype for work from home and thus defying the charity model of disability. Through this paper we want to focus on their expertise in working from home for higher productivity. They suggest that the planning starts with keeping goals, with a structure for deliverable, meeting, and review from superior and peer groups[13]. They recommend keeping a strict schedule and start the work with the most time-sensitive task; the deadline should be finalized and prioritized[13]. People should stretch their arms and legs with an exercise routine to avoid health issues. Working remote needs a well-ventilated separate area with fresh air and greenery around; it helps to keep away the blues. The best practice learned through this experience pave the way toward more understanding and without bias future. It is a fair approach towards remote working for people with or without disabilities in the near future.

---

[15]UN News: Preventing discrimination against people with disabilities in COVID-19 response, https://news.un.org/en/story/2020/03/1059762
[16]Blog Gersh Autism: WHY VIRTUAL LEARNING MIGHT BENEFIT YOUR CHILD ON THE AUTISM SPECTRUM, https://gershautism.com/blog/why-virtual-learning-might-benefit-your-child-on-the-autism-spectrum/
[17]Pennsylvania Assistive Technology Foundation: Social Isolation And Kids With Autism: How Technology Can Help, https://patf.us
[18]https://www.health.govt.nz/our-work/diseases-and-conditions/covid-19-novel-coronavirus/covid-19-resources-and- tools/covid-19-accessible-information/covid-19-new-zealand-sign-language

## 7. Success stories around the world

The policy and implementation structure with relation to people with disability worldwide has provided us with future model of action for any similar situation. We have success stories worldwide providing us with the motivation for some best practices: Paraguay and Panama have evolved systems to ensure relevant information in accessible formats[14]. The New Zealand Ministry of Health has a segment of its website dedicated to information in convenient formats, including sign language[18]. In Argentina, support persons are exempted from movement and physical distancing restrictions to support persons with disabilities[19]. In Colombia, community support networks have volunteers recruited to help people with disabilities, older persons with their groceries and other purchases[14]. In Panama, to reduce infection risks, they have special opening hours for persons with disabilities and their assistants for essential purchases[14]. The United Kingdom and Northern Ireland have relaxed initially strict confinement rules and proposed exemptions to allow autistic persons to go outside [14]. Given priority in the vaccination drive for People with disabilities is a welcoming move toward inclusion, provided disability should not be considered as comorbidity. People with disability should be on the priority list of high risk group irrespective of age and type of disability[20].

## 10. Conclusion

The use of assistive technology justifies the successful implementation of medical model of disability. The use of mobile app to provide the aid through volunteers, accessible health information, ease of access to skilled health care professional are some aspect accounting for the successful accomplishment of social model of disability. Altogether the optimistic approach towards solving the real life issues indicates towards the fulfillment of affirmative model of disability.

The adverse impact of pandemic or any other similar situation can be minimized through proper planning and management well before the emergencies. Due to the Limited availability of disaggregated data, the surveillance system cannot provide a real impact on people with disabilities [6]. This paper ensures to raise awareness about people with disabilities residing in institutions, residential establishments during any situation like pandemic to have access to relevant prevention and response measures; Identification of adults, children with disabilities who require targeted support and information[6]. To create awareness against prejudices, stigma, and misconceptions to remove the discrimination for people with disabilities[6]. We target to raise awareness among policy makers and administration regarding the pandemic response plans that may disproportionally impact persons with disabilities.

All public information, crisis response measures, health and social protection interventions, emergency responses, and recovery programs must be inclusive, accessible for all, and should not discriminate against persons with disabilities [6]. The targeted social protection measures, securing the continuity of home support services and monitoring disability-inclusive humanitarian responses, can play a crucial role in providing essential support [11]. Social care Institution, Nursing Home has become a hotspot of the corona pandemic. Therefore, technological intervention and allowing people with disabilities to live an independent life can save them from transmitting the fatal virus.

---

[19]http://servicios.infoleg.gob.ar/infolegInternet/anexos/335000-339999/335741/norma.htm
[20]https://indianexpress.com/article/india/covid-19-vaccination-scrap-age-bar-for-the-disabled-include-all-categories-of-disabilities-7211229/